\documentclass[preprint]{aastex}

\slugcomment{To appear in the Astrophysical Journal} \shorttitle{Accelerated
Electrons in Cas A } \shortauthors{Laming}
\begin{document}
\title{Accelerated Electrons in Cassiopeia A: An Explanation for the Hard
X-Ray Tail}

\author{J. Martin Laming}
\affil{E. O. Hulburt Center for Space Research, Naval Research Laboratory,
       Code 7674L, Washington DC 20375}
\email{jlaming@ssd5.nrl.navy.mil}

\begin{abstract}
We propose a model for the hard X-ray ($> 10$ keV) emission observed from the
supernova remnant Cas A. Lower hybrid waves are generated in strong (mG)
magnetic fields, generally believed to reside in this remnant, by shocks reflected
from density inhomogeneities. These then
accelerate electrons to energies of several tens of keV.
Around 4\% of the x-ray emitting plasma electrons need to
be in this accelerated distribution, which extends up to electron velocities
of order the electron Alfv\'en speed, and is directed along magnetic field
lines. Bremsstrahlung from these electrons produces the observed hard x-ray emission.
Such waves and accelerated electrons have been observed in situ at
Comet Halley, and we discuss the viability of the extrapolation from this case
to the parameters relevant to Cas A.
\end{abstract}



\section{Introduction}
A number of supernova remnants have recently been recognized as likely
acceleration sites for cosmic ray electrons. SN 1006 \citep{koyama95}
and G347.3-0.5 \citep{slane99,koyama97} both show spectra with continua so
strong in the spectral region around 1 keV that no line emission is detectable with
e.g. the Advanced Satellite for Cosmology and Astrophysics (ASCA).
\citet{laming98} has shown for SN 1006 that any explanation based on
non-thermal bremsstrahlung fails. It must either produce line emission
that should be observable, or in the most extreme case of a shock propagating
through a pure carbon plasma (which will produce no lines in the bandpass of
instruments such as
ASCA), the amount of carbon required to produce the observed intensity is
well in excess of that expected for a Type Ia supernova, and indeed exceeds the
Chandrasekhar limit by large factors. Thus synchrotron radiation from cosmic
ray electrons with energies of order 100 TeV is left as
the only viable interpretation. This conclusion is essentially confirmed by the detection
of TeV $\gamma $-rays from SN 1006 by \citet{tanimori98}.

A second class of cosmic ray accelerator candidates do show the line emission
expected, but also have a hard X-ray ``tail'' to their X-ray spectra. Members
of this class include Cas A \citep{allen97,favata97}, IC 443 \citep{keohane97},
and possibly RCW 86 \citep{allen98}. While less compelling than the first class,
no other mechanism has been postulated to date that could explain the
approximate power law behaviour of the emission, seen in Cas A up to energies
in excess of 100 keV \citep{allen97,the96}. It is the purpose of this paper to
investigate such a possibility, where collective motions of the plasma (specifically
lower hybrid waves) excited by ions reflected from shock front decay accelerating electrons
into a non-Maxwellian distribution function. We begin by reviewing the simplest models
for collisionless shocks. Section 2 reviews the theory of electron energization by
lower hybrid waves,
which has received recent attention in connection with cometary X-ray emission
\citep{bingham97,shapiro99}. Section 3 describes the application of these
results to Cas A, while section 4 discusses the assumptions underlying this
approach, and some of the consequences for our understanding of Cas A that such
a model, if correct, would yield.

Assuming that the gas pressure $>>$ magnetic pressure in the
downstream plasma, the temperature $T_a$ reached by particle
species $a$ with mass $m_a$ is given by conservation of energy,
momentum and particles across the shock discontinuity by
\begin{equation}\label{eqn1.1}
k_BT_a={3\over 16}m_av_s^2
\end{equation}
for shock velocity $v_s$. Thus in the limit of true collisionless
plasma, the particles will have temperature proportional to their
masses, i.e. protons will have a temperature $m_i/m_e = 1836$
times that of the electrons. Obviously in realistic conditions
Coulomb equilibration will take place, with rate
\begin{equation}\label{eqn2.2}
{d\Delta T\over dn_et}=-0.13{Z\over A}{\Delta T\over T_e^{3/2}}
\end{equation}
in c.g.s. units, where $\Delta T=T_i-T_e$, and the ions have
charge $Z$ and atomic mass $A$. So long as $T_i>>T_e$ we can
integrate to get
\begin{equation}\label{eqn1.3}
T_e=\left({5\over 2}\times 0.13{Z\over A}T_in_et\right)^{2/5}.
\end{equation}
Averaging this temperature over the life of the shock modifies
$T_e$ by an extra factor 5/7. Such a model, assuming values of $Z$ in the range
8-26 (corresponding to oxygen and its burning products) adequately explains the
observed temperatures of shocked ejecta and circumstellar medium
(CSM), 1.25 and 3.8 keV respectively for $n_eT\sim 10^{11}$
cm$^{-3}$s in each case in Cas A \citep{favata97}, but cannot
explain the hard x-ray continuum extending out to 100 keV.

\section{The Electron Distribution Function from Lower Hybrid Wave Damping}
From observations of radio synchrotron emission Cas A is believed to have
relatively high magnetic fields, perhaps
up to the equipartition value of order 1 mG \citep{gull75,keohane98}. A total energy in
cosmic ray electrons and magnetic field of $\sim 10^{49}$ erg is required
\citep[c.f.][]{longair94}, which for
maximum synchrotron luminosity must be equipartitioned between the two. If distributed
uniformly throughout the Cas A shell, this evaluates to a magnetic field of 0.5 mG.
Thus lower hybrid waves excited by cross field ion motions can accelerate
electrons into a high energy tail onto the
otherwise thermal Maxwellian distribution function.
These are electrostatic ion waves that
occur when the electrons are magnetized, which is likely to be the case at so-called
quasi-perpendicular shocks. Due to the the shock
compression of the magnetic field component perpendicular to the
shock velocity vector, quasi-perpendicular shocks are likely to be
more prevalent than quasi-parallel shocks in Cas A, and in supernova remnants in general,
and will be assumed throughout this paper. Electron velocities at least up to
the electron Alfv\'en velocity, $B/\sqrt{4\pi n_em_e}$ where $B$ is the
magnetic field strength, and $n_e$ and $m_e$ are the electron density and mass
respectively, are possible. This corresponds to energies of order 100 keV for
Cas A, making it an attractive mechanism to explore.

Electrons accelerated by these waves emit bremsstrahlung
radiation, and have been recently discussed as a mechanism for cometary x-ray emission.
Neutral atoms or molecules outgassing from a comet nucleus can be
photoionized by solar UV radiation. In the interaction of these
stationary (in the comet rest frame) ions with the ambient solar
wind, a modified two-stream instability develops that in the case of
magnetic field perpendicular to the relative comet-solar wind
velocity generates lower hybrid waves \citep{bingham97,shapiro99}.

We will apply a similar model to Cas A, following in part the
analytic treatments of \citet{vaisberg83}, \cite{krasno85}, and \citet{begelman88}, where
ions reflected from shock fronts in Cas A move back upstream through the preshock plasma
generating a similar modified two-stream instability. Such reflected ions and associated wave
generation are common features in in situ observations of solar system shocks
\citep[c.f.][]{lengyel97}, and of models of electron energization at collisionless shocks
\citep[c.f.][]{cargill88,tokar86,mcclements97,vaisberg83,krasno85}.
\citet{vaisberg83} give an analytic form for the electron distribution function following
lower hybrid wave energization, which agrees well with the particle-in-cell
simulations performed by \citet{shapiro99}.
The mathematical development of the theory for these waves
is summarized in Appendix A. It is shown that for both reactive and kinetic forms of the
instability,
wave growth is strongest for wavevectors with a small parallel
component $k_{||}/k\le\omega _{pi}/\omega _{pe}$. Hence
the wave can simultaneously be in resonance with ions moving
across and electrons moving parallel to the magnetic field, since the phase velocities are
much larger along the magnetic field than across it. This
facilitates energy transfer between electrons and ions.
The resulting electron distribution function parallel to the magnetic field is also
derived in Appendix A.

However the wave group
velocity away from the shock front is lower than the shock velocity itself for
$k_{||}/k<\omega _{pi}/\omega _{pe}$. Only at $k_{||}/k=\omega _{pi}/\omega _{pe}$
or greater can the waves stay ahead of the shock. This is important because the growth rates,
particularly for the kinetic form of the instability, are not large, and for smaller $k_{||}$
the waves are likely to be overrun by the shock before significant growth can occur. Hence
in the following we assume that in the preshock region, a significant level of wave activity
exists only for $k_{||}/k=\omega _{pi}/\omega _{pe}$. With this simplification the
accelerated electron distribution function is then given by equation (A17) with $\cos\theta
=\omega _{pi}/\omega _{pe}$
\begin{equation}\label{eqn2.1}
f_e\left(v_{||}\right)={n_i^{\prime}\over\sqrt{2\pi }k_BT/m_i}
\exp\left(-{1\over2}\right){\omega _{pi}^2\over\omega _{pe}^2}
\left(v_m-v_{||} +{\left(v_m^3-v_{||}^3\right)\over 3v_{Ae}^2}\right)
\end{equation}
where $v_{Ae} = v_{Ai}\sqrt{m_in_i/m_en_e}$ is the electron
Alfv\'en velocity, ($v_{Ai}=B/\sqrt{4\pi n_im_i}$ being the usual Alfv\'en velocity),
$n_i^{\prime}$ is the density of shock reflected ions,
and $v_m$ is a constant of integration, denoting the maximum electron velocity.
In the quasi-linear theory used in Appendix A, the accelerated electron density,
$n_e^{\prime}$, or $v_m$ remains a single free
parameter. They are related by
\begin{equation}\label{eqn2.4}
n_e^{\prime}={n_i^{\prime}\over\sqrt{2\pi }k_BT/m_i}
\exp\left(-{1\over 2}\right){\omega _{pi}^2\over\omega _{pe}^2}
\left(v_m^2 +{v_m^4\over 2v_{Ae}^2}\right).
\end{equation}
where an extra factor of two has been included to include the electron velocity range
$-v_m\rightarrow +v_m$.

For a reflected ion population $\sim 25$\% of the preshock ion density
\citep[c.f.][and references therein]{cargill88,mcclements97}, we get
\begin{equation}
{n_e^{\prime}\over n_e}=0.06\left({v_m^2\over v_{Ae}^2} +{v_m^4\over 2v_{Ae}^4}\right)
{m_e\over m_i}{v_{Ae}^2\over v_{ti}^2}.
\end{equation}
Anticipating from the following section that $n_e^{\prime}/n_e\simeq 4$\%,
$v_{Ae}\sim 26v_s$ and $v_{ti}=0.53v_s$ (assuming complete equipartition of magnetic and
ion thermal energy, but insignificant electron-ion equilibration) gives
$v_m=v_{Ae}\sqrt{\sqrt{1+A} -1}$ where $A$ is the atomic mass of the ions. Hence
$v_m\sim v_{Ae}$, slowly increasing with the ion mass.

\section{Application to Cassiopeia A}
We compute model bremsstrahlung spectra using the distribution function
given by equation (4) above for the electron velocity parallel to the
magnetic field, representing the electron velocity perpendicular to the
magnetic field as Maxwellians. The temperature for these perpendicular
components is taken to be the same as that fitted to the thermal bremsstrahlung
emitted at energies less than 10 keV.  A fit to the
BeppoSAX Medium Energy Concentrator \citep[MECS][]{boella97} spectrum from
1997 November 26
determined this temperature at 3.3 keV. This fit is shown in Figure 1.
Other workers find temperatures in the range 2.9-4.2 keV
\citep{allen97,favata97,vink96} from ASCA and BeppoSAX data.

The non-relativistic
bremsstrahlung cross section $d\sigma $
for photon emission with energy in the range $E\rightarrow E+dE$
is given by \citep{berestetskii82}
\begin{equation}
d\sigma = -{d\over dz}\left|_2F_1\left({iZ\over v_f},{iZ\over v_i},1,z\right)
\right| ^2
{64\pi ^2\over 3}Z^2\alpha r_0^2{m_e^2c^2\over\left(p_i-p_f\right)^2}
{p_f\over p_i}{1\over\left(1-\exp\left(-2\pi Z/v_f\right)\right)\left(\exp
\left(2\pi Z/v_i\right)-1\right)}{dE\over E}
\end{equation}
where $p_i$, $p_f$, $v_i$, $v_f$ are initial and final electron momenta and
velocities, $Z$ is the nuclear charge, $\alpha$ is the fine structure
constant, and $r_0$ is the classical electron radius.
Using standard formulae for manipulating the hypergeometric function
${}_2F_1$ and its derivative
in the limit of low $Z/v_i$ and $Z/v_f$ a simpler form suitable
for evaluation of the double integral to compute the emission rates can be
found;
\begin{equation}
d\sigma ={64\pi^2\over 3}Z^2\alpha r_0^2{c^2\over v_i^2}
{Z/v_f\over\left(1-\exp\left(-2\pi Z/v_f\right)\right)}
{Z/v_i\over\left(\exp\left(2\pi Z/v_i\right)-1\right)}
\log\left(p_i+p_f\over p_i-p_f\right){dE\over E}.
\end{equation}
In checks against bremsstrahlung spectra computed with the full formula given
by equation (7) \citep{sutherland98}, the simpler equation
(8) was found agree at the few percent level for Maxwellian electron
distributions, for temperatures appropriate to Cas A for target charges up to 8
(i.e. a fully ionized oxygen plasma). Cas A has very little hydrogen.
Oxygen is probably the most abundant element \citep{vink96}.

In Figures 2 and 3 we plot spectra from the BeppoSAX MECS and the
higher photon energy
Phoswich Detector System \citep[PDS][]{frontera97} instruments, together with model spectra.
The softest model spectrum in Figure 2 is a pure thermal
bremsstrahlung spectra at 3.3 keV in fully ionized oxygen. The emission measure
for this spectrum is $n_en_OV=3.05\times 10^{57}$ cm$^{-3}$ for an assumed
distance to Cas A of 3.4 kpc \citep{reed95}. The successively harder spectra
have been computed by taking 4\% of these thermal electrons and putting them
into a lower hybrid wave energized distribution function, for electron Alfv\'en
velocities of 32, 40, and 48 atomic units (in atomic units, $e=\hbar =m_e
=1$, so the atomic unit of velocity is $2.188\times 10^8$ cm s$^{-1}$), with $v_m=1.77v_{Ae}$.
Figure 3 shows the same data with the pure thermal bremsstrahlung and non-thermal bremsstrahlung
models for $v_{Ae}$ of 56, 68, and 80 atomic units for a pure He plasma,
with $v_m=1.11v_{Ae}$. In both cases, the curves with $v_m\sim 70$ atomic units appear to
give the best match to the data up to energies $\sim 50$ keV.

\section{Discussion}
\subsection{The Morphology of Cas A}
One immediate problem is that the mG magnetic fields are present only inside
the Cas A remnant (presumably amplified by Rayleigh-Taylor instabilities at the contact
discontinuity), whereas most shock models require such fields to be present
{\em ahead} of the shock (to be discussed in more detail below). However Cas A
is an extremely inhomogeneous supernova remnant. It has numerous knots of optical emission
from low charge states, which must be considerably denser than the surrounding plasma
so that the recombination and radiative cooling times are sufficiently short.
The blast wave travelling
through the circumstellar material will to split into transmitted and
reflected shocks upon meeting these inhomogeneities
\citep[see][for fuller discussion]{sgro75,borkowski92}. Indeed exactly this
happens in the simulations of \citet{borkowski96}, giving rise to the
possibility that shock waves are propagating throughout the X-ray and radio
emitting shell. Since the plasma has already been heated by the blast wave or reverse shock
these shocks will most likely be of relatively low Mach number (but still supercritical; see
subsection 4.5 below).
Another possibility for inhomogeneous or clumpy SNR is that the
Rayleigh-Taylor fingers from the unstable contact discontinuity can propagate
outwards and penetrate through the forward shock \citep{jun96}.
The by-now famous Chandra x-ray images \citep{hwang00}
suggest that both scenarios could indeed be true.
The forward shock is visible as a faint ring around
the outside of the bright x-ray shell, which has its origin in shocked ejecta. The
boundary between faint and bright emission is the contact discontinuity, which has
no obvious penetration through the blast wave, except in the ``jet'' region in the NE
quadrant. However the apparent
``inversion'' of the presumed pre-supernova element stratification \citep{hughes00,hwang00},
with Fe being found in outer regions of the eastern part of the
remnant with respect to lighter elements like Si, suggests a role for the Raleigh-Taylor
instability, and
makes it more likely that these regions have come into contact with strong magnetic fields
at the contact discontinuity. Lower hybrid electron acceleration is also more likely in
the heaviest element plasmas, (i.e. the Fe ejecta), for reasons discussed in the next
subsection.

\subsection{The Thermal Electron Temperature}
The lifetime of accelerated electrons against Coulomb collisions with
the thermal background is $t\simeq 5\times 10^6\left(E/1{\rm ~keV}\right)^{3/2}$ s
which evaluates to 14 - 150 years for electron energies of 20 - 100 keV. Hence electron
acceleration must be ongoing, or have ceased in the very recent past, though this last
possibility seems unlikely since the hard x-ray emission from Cas A has been known since
the earliest days of x-ray astronomy \citep{gorenstein70}.

In section A.4 of the Appendix it is shown that an appropriate criterion for the presence
of lower hybrid waves is that the electron gyroradius should be less than the wavelength,
which can be expressed as $T_e/T_i < \Omega _e^2/Z/\omega _{pe}^2$. In a magnetic field of
1 mG and electron density of 10 cm$^{-3}$ this evaluates to $T_e<<3\times 10^6\left(v_s/4000
{\rm km~s}^{-1}\right)^2$.
The shock velocity has been determined from proper motion studies. For the same
assumed distance as before, radio data give an expansion velocity of $\sim $
2000 km s$^{-1}$ \citep{anderson95}, while X-ray studies give velocities of
$\sim$ 3500 km s$^{-1}$ for the bright ring at a radius of 110'', and $\sim$
5200 km s$^{-1}$ inferred for the blast wave at a radius of 160''
\citep{vink98}, all of which require $T_e$ significantly lower than that observed.
As discussed in section A.4, an alternative condition on
the electron Landau damping rate requires $k_BT<<m_e\Omega _{LH}^2/k_{||}^2$, which is
the same as above for $k_{||}/k\simeq\omega _{pi}/\omega _{pe}$, but less restrictive
for smaller $k_{||}$. The first criterion is also not met in the plasma around the nucleus
of Comet Halley where lower hybrid waves and accelerated electrons were directly detected
in situ \citep{gringauz86,klimov86}. They observed $\Omega _{LH}\sim 100$ rad s$^{-1}$,
and $n_e\simeq 300$ cm$^{-3}$ giving $\Omega _e^2/\omega _{pe}^2\sim2\times 10^{-5}$, while
$T_e\sim 2\times 10^5$ K and $T_i\sim 10^7$ K.

One plausible solution to this dilemma is that in the region of wave generation, the
electron temperature is in fact lower than the estimates and observations above would
indicate. This is possible if the electrons are not energized on passage through the
shock due to being bound to ions, only becoming ionized some time later on either by
collisions in Cas A or by photoionization in Comet Halley. This is demonstrated in
calculations of the Cas A reverse shock propagating through clumps of Fe ejecta by
\citet{mochizuki99}, where temperatures satisfying the condition on the electron
gyroradius are found. Thus for Cas A the hard x-ray
emission is more likely to be associated with regions of shocked ejecta where many electrons
are still being ionized from bound states. In order that this putative ionizing region in
Cas A should produce enough bremsstrahlung emission to be visible against the thermal
continuum, a higher background ion charge might be necessary. This is also more likely
if the thermal bremsstrahlung continuum with $T_e=4\times 10^7$ K comes from the shocked
circumstellar medium, composed of He, N, and possibly O, while the non-thermal component
comes from the ejecta, composed of O and heavier elements. Thus the local fraction of
accelerated electrons may be different from the 4\% determined above by matching to the
BeppoSAX MECS/PDS spectra. However the maximum accelerated electron velocity
$v_m\propto\left(n_e^{\prime}/n_e\right)^{1/4}$ is relatively insensitive to this, making
large changes in the bremsstrahlung spectrum from that calculated before unlikely. The
temperature assumed above for the perpendicular components of the accelerated electron
distribution will also change. Again this will not significantly alter the non-thermal
bremsstrahlung spectrum.

Another reason for spatially separating the thermal and non-thermal bremsstrahlung emitting
regions is that over the 300 years or so of the remnant's existence, collisions between the
accelerated and thermal electrons would be likely to have raised the thermal electron
temperature to a value well in excess of that observed if they were cospatial.

\subsection{The Electron Alfv\'en Speed}
Another crucial parameter in determining the visibility of non-thermal
bremsstrahlung is the electron Alfv\'en velocity. From observations of radio
synchrotron emission, Cas A is believed to have a large magnetic field $\sim
1$~mG, generated by turbulence associated with Rayleigh-Taylor instabilities at
the contact discontinuity between the shocked ejecta and circumstellar material
\citep[c.f][]{gull75,keohane98}.
Assuming such instabilities generate magnetic fields up
to equipartition, we may write
\begin{equation}
{B^2\over 8\pi} = {1\over 2}\left(n_ik_BT_i +n_ek_BT_e\right)
\end{equation}
where $T_i$ and $T_e$ are given by the shock jump conditions (equation 1).
The resulting electron Alfv\'en velocity thus becomes
\begin{equation}\label{eqn12}
v_{Ae}=\sqrt{B^2\over 4\pi n_em_e}=v_s\sqrt{3m_p\over 8m_e}=26v_s
\end{equation}
for fully ionized plasmas where nuclear mass $Am_p$ is such that $A=2Z$.
The X-ray blast wave velocity (discussed above) gives $v_{Ae}\simeq 60$ atomic
units according to equation (\ref{eqn12}), with correspondingly lower values from the other
observations. This is more than adequate to give a sufficiently high $v_{Ae}$ for the
non-thermal bremsstrahlung to originate in oxygen or heavier element plasma. Hard x-rays
from a helium plasma require higher $v_{Ae}$, still consistent with the observed x-ray
blast wave, but not with the lower inferred shock velocities.

\subsection{High Energy Behaviour of the Continuum and $^{44}$Ti}
The highest energy data points (above 100
keV) may still possibly require the existence of cosmic rays. As can be seen in Figures 2 and
3, $v_{Ae}$ or $v_m$ cannot be arbitrarily increased to reproduce this emission without
overestimating the non-thermal continuum at lower energies. However in
principle relativistic electron energies are possible from lower hybrid wave acceleration
\citep{mcclements97,vaisberg83}. The maximum cutoff in our approach stems from taking only
waves with $k_{||}/k=\omega _{pi}/\omega _{pe}$, which satisfies the
requirement that the perpendicular group velocity of the waves be greater than the
shock velocity, to allow sufficient time for wave growth. With stronger reactive instabilities,
waves with $k_{||}\rightarrow 0$ may exist at sufficient intensity to make
electron acceleration by lower-hybrid waves a plausible mechanism for solving the
so-called electron injection problem \citep{mcclements97}.
This arises because first order Fermi acceleration at
quasi-perpendicular shock fronts is only effective for electrons of mildly relativistic or
higher energies \citep{levinson96}, and
so some other ``injection'' mechanism is required.

The question of the spectral behaviour at high energies is further confused slightly by the
discrepancy in fluxes observed above 100 keV by BeppoSAX and OSSE. Compared with the data from
OSSE presented in Figure 2 of \citet{allen97}, the BeppoSAX PDS fluxes are up to an order of
magnitude higher above 100 keV. Cas A is unlikely to be a variable source at these (or
any other) energies, and so this problem presumably lies with the calibration of one or other
of the instruments.

Another issue complicating the comparison of continuum spectra, but of
supreme interest in its own right, is the possibility of detecting $\gamma$-ray line emission
from the nucleus $^{44}$Sc at energies of 67.9 and 78.4 keV. $^{44}$Sc is a decay product
of $^{44}$Ti, which is one of the most important radioactive elements produced in the
$\alpha $-rich freeze out of core collapse supernovae, its abundance also being sensitive
to the position of the mass cut
\citep[c.f][for further discussion]{vink00}. The break in the lower-hybrid wave accelerated
electron spectrum occurs at energies very close to where the $^{44}$Sc lines should be
present, and the substitution of the power law in the fit to the BeppoSAX PDS data by
\citet{vink00} with the non-thermal bremsstrahlung spectrum modeled in this paper might
lead to interesting results for the flux in these lines.

\subsection{Properties of Collisionless Shocks}
So far we have invoked the excitation of lower hybrid waves ahead of a shock
front, and modeled the electron distribution function on the assumption that a
quasi-steady state can be reached in the shock rest frame. We will discuss
these assumptions in more detail here.

In a number of models, various types of plasma waves are excited by ions that
are reflected from the shock front and travel back upstream. This generally arises in
perpendicular shocks due to the cross shock ambipolar electric field in the magnetic
field overshoot region. The simulation of
\citet{cargill88} has reflected ions exciting electron plasma waves and then
ion acoustic waves to provide eventual electron heating to around 20\% of the
postshock ion temperature. \citet{mcclements97} looked at the excitation of
lower hybrid waves by ion gyrating in the magnetic field upstream of a fast
shock, with a view to modeling the injection mechanism for the Fermi
acceleration of electrons to cosmic ray energies at perpendicular shocks. A
certain amount of support for these models comes from observations of shock
waves in the solar wind where electron plasma and ion acoustic waves are seen
ahead of shock fronts \citep{lengyel97}, and lower hybrid waves
\citep{klimov86} and their associated electron distribution function
\citep{gringauz86} have also been observed at comet Halley, with
\citet{shapiro99} recently demonstrating by particle-in-cell simulations that
the observed wave activity is indeed consistent with the observed electron
distribution function.

An important point concerns the nature of the two-stream instability. All of
the models discussed above assume the ions reflected from the shock to be
essentially monoenergetic, thus giving rise to fast growing reactive
instabilities. At perpendicular shocks this is necessary since the reflected
ions only move back upstream for about an ion gyroradius before begin swept
back behind the shock front, thus allowing only a finite amount of time for the
instability to develop. This monoenergetic property of the reflected ions
surprisingly appears to persist in shock simulations to high Mach numbers,
where the shocks themselves should be highly turbulent
\citep{cargill88,tokar86}. Both sets of authors comment that the ion reflection
in their simulations becomes ``bursty'', giving a time average of $\sim 25$\%;
not too different from the fraction of preshock ions observed to be reflected
at lower Mach number (less turbulent) shocks in the solar wind or laboratory
experiments. \citet{woods87} emphasises that such ion reflection is
characteristic of a very thin shock transition. For thicker shocks, the ions
will gyrate in the magnetic field before reflection to varying degrees, giving
rise to a reflected ion distribution still monoenergetic, but having a finite
angular spread. \citet{woods87} terms this a transition from ``specular'' to
``diffuse'' reflection. \citet{gedalin96} has considered this distinction in
more quantitative detail.
Diffusely reflected ions will produce slower growing
instabilities ahead of the shock, and hence produce less electron heating, and
have been suggested as a possible reason by \citet{laming98} for the low amount
of collisionless electron heating observed at fast ($\sim 2600$ km s$^{-1}$)
shocks in the NW region of SN 1006 by \citet{laming96}.

We have argued above that based on our knowledge of the properties of Cas A, the
shocks producing the reflected ions and lower hybrid waves are likely to be of relatively
low Mach number, since they are reflected from density inhomogeneities and are
propagating through plasma that has already been heated
by the forward or reverse shocks. Hence a ``bursty'' ion reflection is unlikely, and the
assumption of steady state conditions embodied in equation (A15) remains valid.
The sound speed in plasma shocked by the blast wave with speed $v_s$ and with no electron-ion
equilibration is $\sqrt{5k_BT_i/3m_i}=\sqrt{5/16}v_s$. This plasma continues expanding at
speed $0.75v_s$, and a shock reflected from a density inhomogeneity will hit it with relative
velocity $\sqrt{15A_r/16/\left(4-A_r\right)}v_s$ (i.e. a Mach number of
$\sqrt{3A_r/\left(4-A_r\right)}$), where $A_r$ is related to the
density contrast, $A$, by $A=3A_r\left(4A_r-1\right)/\left(\sqrt{3A_r\left(4-A_r\right)}
-\sqrt{5}\left(A_r-1\right)\right)$ \citep{sgro75}.
For $A\rightarrow\infty$, $A_r=2.5$ and the maximum
Mach number of the reflected shock is $\sqrt{5}$.
In order that these secondary shocks
propagating back through the shell are supercritical (i.e. that they reflect ions to form
a precursor that excites the lower-hybrid waves), they must have Mach number greater than
the so-called first critical Mach number. \citet{edmiston84} in a survey of first critical
Mach numbers for various MHD shocks find a value of 1.7 for perpendicular shocks in
plasmas with $\gamma$ (ratio of specific heats) of 5/3 and $\beta$
(gas pressure/magnetic pressure) of $\sim 1$.
This requires $A_r\simeq 2$ and a density contrast of $\sim 30$, with lower values required
for higher $\beta$. For reference the electron density in the shocked x-ray emitting
plasma is in the range 10-15 cm$^{-3}$ \citep{vink96,favata97}, while that in the shocked
optically emitting knots determined from S II line ratios
is $\sim 10^3-10^4$ cm$^{-3}$ \citep{chevalier78,reed95}. The initial
mass density contrast will be increased from the simple ratio of these electron densities
since the x-ray emitting plasma is more highly ionized than the optical knots, by a factor
of $\sim 10$. However the density of the optical knots may also have been enhanced over that
due to shock compression by radiative cooling. \citet{chevalier78} report densities of
$< 100$ cm$^{-3}$ from diffuse (presumably unshocked) S II, which given the difference in
ionization state between this and the x-ray emitting plasma still gives sufficient density
contrast to reflect supercritical shocks back into the shell.

Whether the ion reflection is diffuse or specular is a harder question
to answer. We have assumed a kinetic form for the instability, since this is likely to be
more realistic, and also gives a simpler derivation of
an analytic form for the electron distribution function. In order to get sufficient
time for wave growth to occur ahead of the shock, we are limited to waves with parallel
components of wavevector $k_{||}/k = \omega _{pi}/\omega _{pe}$ or greater, in order to have
group velocities away from the shock comparable to the shock velocity itself. Specularly
reflected ions give rise to much faster growing instabilities, where waves with smaller
$k_{||}$ may grow to significant intensities before being overrun by the shock to allow
harder electron spectra to be produced, and might conceivably produce a better match to
the BeppoSAX MECS/PDS data. This issue of specular versus diffuse ion
reflection and the reality of the electron distribution function discussed
here is likely to remain an open question at least until high resolution
X-ray spectra become available.
We note that the electron distribution function produced by the wave
activity should have discernable effects on the line spectrum. Ionization and
excitation rates will be altered by the extra population of high energy
electrons, and modeling of the X-ray spectrum might be the best way to confirm
or refute the existence of such an electron distribution function
\citep[see][for an example]{gabriel91}. Such an
effort is beyond the scope of this paper, and requires high spectral
resolution observations where individual lines can observed.

\section{Conclusions}
We have presented a model based on plasma wave excitation by collisionless
shocks to generate a non-thermal electron distribution, the bremsstrahlung from
which can explain the hard X-ray ``tail'' in the spectrum of Cas A, for very
reasonable model parameters. This is important because the alternative
interpretation, that of X-ray synchrotron emission from cosmic ray electrons
with energies of order $10^{14}$ eV, appears to have been accepted largely on
the basis that it is the {\em only} interpretation available, though some further
observational support for this view is accumulating \citep{vink99,hwang00}. Our model is
based on observation of the relevant waves and energized electrons at Halley's
comet, and analytical and numerical treatments of the process confirming the
consistency of the observed electron distribution
with the observed wave activity. We have assumed that the wave excitation is associated
with shocks which propagate throughout the shell of Cas A, with the waves mainly being
generated in ionizing regions of the ejecta, although other mechanisms could be
viable. For instance, \citet{begelman88} discuss similar waves excited by relative
electron-ion drifts in Advection Dominated Accretion Flows. It is also worthwhile to
point out that electron acceleration by electrostatic waves is essentially instantaneous
on the timescales relevant to supernova remnant evolution, whereas for example
stochastic mechanisms (i.e. second order Fermi acceleration) are not \citep[c.f.][]{blasi00}.

Similar explanations may also be valid for the
supernova remnants RCW 86 and IC 443, for which hard X-ray tails have also been
observed. \citet{chevalier99} has also expressed scepticism that such emission
in IC 443 is due to shock accelerated cosmic ray
synchrotron emission. Our discussion has concentrated on
Cas A simply because of its well known age, distance, shock velocities, and
conspicuous high energy continuum. If proved correct, then this explanation could provide the
clearest evidence to date for equipartition of magnetic and thermal energy
densities, and the precise shape of the non-thermal bremsstrahlung spectrum should be
rich in diagnostic potential for inferring magnetic fields, shock structure and the
likely nature of preshock ion driven instabilities.

Finally we note that our non-thermal bremsstrahlung model is qualitatively
different to those of previous authors \citep{asvarov90,sturner96}. Both
references consider non-thermal bremsstrahlung in X-rays from the same electron
population that produces synchrotron radiation emission, and find that for the
hard X-ray tail to be explained in this way, it is likely that
too much radio emission would
result. \citet{baring00} give a fuller discussion of such scenarios.
Our model here postulates a separate electron population which is
non-relativistic; basically just a
small distortion of the thermal distribution, which need not necessarily be associated with
the cosmic ray electron distribution, and so no
great change in radio emission should result.

\acknowledgments This work has been supported by basic research funds of the
Office of Naval Research, and has also made use of data obtained from the
High Energy Astrophysics Science Archive Research Center (HEASARC), provided
by NASA's Goddard Space Flight Center.
I am grateful to John Raymond for bringing the comet
work to my attention, and to Jacco Vink and Matthew Baring
for helpful discussions.

\appendix
\section{Appendix: Lower Hybrid Waves}
\subsection{Dispersion Relation and Reactive Growth Rate}
For cold magnetized electrons the dielectric tensor is
\begin{equation}
K_{ij}=\left(\matrix{ 1-{\omega _{pe}^2\over\omega ^2-\Omega
_e^2}& i{\Omega _e\over\omega}{\omega _{pe}^2\over\omega ^2-\Omega
_e^2}& 0\cr -i{\Omega _e\over\omega}{\omega _{pe}^2\over\omega
^2-\Omega _e^2}& 1-{\omega _{pe}^2\over\omega ^2-\Omega _e^2}&
0\cr 0 & 0 & 1-\omega _{pe}^2\over\omega ^2\cr}\right)
\end{equation}
where $\omega _{pe}$ and $\Omega _e$ are the electron plasma and
gyrofrequencies respectively. For longitudinal waves we put ${\bf
E} ||{\bf k}$ in the wave equation, take the scalar product with
${\bf k}$ with the magnetic field along the $z$-axis and set the
result equal to zero to derive the dispersion relation;
\begin{equation}
k_iK_{ij}k_j=\left(\matrix{\sin\theta ,
&0,&\cos\theta\cr}\right)K_{ij} \left(\matrix{\sin\theta \cr 0 \cr
\cos\theta \cr}\right) =0
\end{equation}
which gives
\begin{equation}
K^L=1-\left(\omega _{pe}^2\over\omega ^2-\Omega _e^2\right)\sin
^2\theta - {\omega _{pe}^2\over\omega ^2}\cos ^2\theta =0.
\end{equation}
To this we can add a term for thermal non-magnetized ions
$\left(1-\phi\left( \omega/\sqrt{2}kv_i\right)\right)\omega
_{pi}^2/k^2v_i^2$ where $\phi\left(x\right)$ is the usual plasma
dispersion function and $v_i=\sqrt{k_BT_i/m_i}$, the ion thermal
velocity. In the limit $\omega >> kv_i$, $\phi\left(x\right)
\rightarrow 1+1/2x^2+\dots $ and the well known form for the
dielectric tensor for lower hybrid waves results
\begin{equation}\label{eqn1}
 K^L=1+{\omega _{pe}^2\over\Omega _e^2}\sin ^2\theta
-{\omega _{pi}^2\over\omega ^2} -
 {\omega _{pe}^2\over\omega ^2}\cos ^2\theta =0
\end{equation}
where $\omega _{pi}$ is the ion plasma frequency. This yields the
dispersion relation $\omega \simeq\omega _{pe}^2$ as
$\cos\theta\rightarrow 1$ and $\omega\simeq\Omega _{LH}^2$ as
$\sin\theta\rightarrow 1$. $\Omega _{LH}$ is known as the Lower
Hybrid frequency and is equal to $\sqrt{\Omega _e\Omega _i}$, with
$\Omega _i$ being the ion gyrofrequency. We now consider a
further ion population moving with velocity $U$ with respect to
the stationary plasma. The dispersion relation becomes
\begin{equation} \label{eqn2}
 K^L=1+{\omega _{pe}^2\over\Omega _e^2}\sin ^2\theta
-{\omega _{pi}^2\over\omega ^2} -
 {\omega _{pe}^2\over\omega ^2}\cos ^2\theta -{\omega _{pi}^{\prime ~2}\over
\left(\omega - kU\right)^2}=0.
\end{equation}
For $\cos\theta\rightarrow 1$ equation (\ref{eqn2}) takes the same
form of that for the Buneman instability where Langmuir waves are
generated by the ion beam \citep[c.f.][pp 113]{sturrock94}. For $\cos\theta\rightarrow 0$ a quartic
equation similar to that for the Buneman instability results, but with the
replacements $\omega _{pe}\rightarrow\Omega _{LH}$ and $\omega
_{pi}\rightarrow\Omega _{LH}\omega _{pi}^{\prime}/\omega _{pi}$,
giving a growth rate
\begin{equation}\label{eqn2a}
\gamma = {\sqrt{3}\over 2^{4/3}}\Omega
_{LH}\left(n_i^{\prime}\over n_i\right)^{1/3}.
\end{equation}
Moving away from $\cos\theta =0$, we put $\omega =\Omega _{LH}$ in
the parallel electron term and the ion term to get
\begin{equation}
\left(\omega -kU\right)^2={\omega _{pi}^{2~\prime}\over 1+{\omega
_{pe}^2\over \Omega _e^2}\sin ^2\theta -{\omega _{pe}^2\over\Omega
_{LH}^2}\cos ^2\theta -{\omega _{pi}^2\over\Omega _{LH}^2}}.
\end{equation}
With $kU\sim\Omega _{LH}$ and $\omega _{pe}>>\Omega _e$, the
growth rate $\gamma$ is
\begin{equation}
\omega - \Omega _{LH}\simeq\pm i\gamma\simeq\pm
i\left(n_i^{\prime}\over n_i\right)^{1/2} {\Omega
_i\over\cos\theta }.
\end{equation}
Hence for $\cos\theta {<\atop\sim}{\omega _{pi}/\omega _{pe}}$,
the growth rate is of order $\Omega _{LH}$, falling off as $\Omega
_i/\cos\theta $ for larger $\cos\theta $.

As discussed by \citet{begelman88} an electromagnetic correction
is required in the parallel electron term of $K^L$. Since waves
propagate within $\pm\omega _{pi}/\omega _{pe}$ of the
perpendicular direction to the magnetic field, the wave phase
speed parallel to the magnetic field can be very large,
approaching or exceeding the speed of light. This correction can
be written $\omega _{pe}^2\rightarrow\omega _{pe}^2/\left(1+ \omega
_{pe}^2/k^2c^2\right)$ \citep[p 28]{begelman88,melrose86}. We also retain further terms in the
expansion; $1/\left(\omega ^2-\Omega _e^2\right)\simeq -1/\Omega _e^2\left(1+
\omega ^2/\Omega _e^2+\dots\right)$, with the last term in parentheses
$\simeq\Omega _{LH}^2/\Omega _e^2 \simeq\omega _{pe}^2v_{Ai}^2/\Omega _e^2c^2
\sim\omega _{pe}^2/k^2c^2\cdot v_{\perp}^2/v_e^2$. Here $v_{\perp}^2=v_{Ai}^2+v_e^2$ is the
electron velocity perpendicular to the magnetic field, with $v_{Ai}$ the Alfv\'en velocity,
$v_e$ the electron thermal velocity, and
$c$ the speed of light. Taking $k\sim\Omega _e/v_{\perp}$, with $k_{||}$ and $k_{\perp}$ being
wavevectors parallel and perpendicular to the magnetic field, applying these two corrections
yields \citep[see e.g.][]{vaisberg83,mcclements97}
\begin{equation}\label{eqn3}
 K^L=1+{\omega _{pe}^2\over\Omega _e^2}{k_{\perp}^2\over k^2}\left(1+{\omega
 _{pe}^2\over k_{\perp}^2c^2}{1\over 1+v_e^2/v_{Ai}^2}\right) -{\omega _{pi}^2\over\omega ^2} -
 {\omega _{pe}^2\over\omega ^2}{k_{||}^2\over k^2}\left(1+{\omega
 _{pe}^2/k^2c^2}\right)^{-1}.
\end{equation}
\citet{mcclements97} discuss reactive growth rates derived using
similar dispersion relations for various reflected ion
distributions.

\subsection{Kinetic Growth Rate and Accelerated Electron Distribution}
In the presence of suprathermal electrons and ions, two extra
terms are added to the expression for $K^L$,
\begin{equation}\label{eqn4}
+{\omega ^{\prime ~2}_{pe}\over n^{\prime}_ek^2} \int
{\vec{k}\cdot\vec{v_e}\over\omega} {\vec{k} \over \omega
-\vec{k}\cdot\vec{v_e}}\cdot{\partial f_e^{\prime}\over\partial\vec{v_e}}
d^3\vec{v_e} +{\omega ^{\prime ~2}_{pi}\over n^{\prime}_ik^2} \int
{\vec{k}\cdot\vec{v_i}\over\omega} {\vec{k} \over \omega
-\vec{k}\cdot\vec{v_i}}\cdot{\partial f_i^{\prime}\over\partial\vec{v_i}}
d^3\vec{v_i},
\end{equation}
where the primes denote quantities with respect to the
suprathermal particle distributions, $n^{\prime}_e$ and
$n^{\prime}_i$ being the suprathermal electron and ion densities,
respectively. The overall growth rate for lower-hybrid waves is
found by setting $\omega\rightarrow\Omega _{LH} +i\gamma$ and
taking the imaginary parts, using the Landau prescription for the
terms in equation (\ref{eqn4}). This produces a growth rate
$\gamma$;
\begin{eqnarray}\label{eqn5}
\nonumber 2\gamma =&{\pi\Omega _{LH}^3\omega ^{\prime
~2}_{pi}\over k^2n^{\prime}_i}\left[ \omega _{pi}^2+\omega
_{pe}^2{k_{||}^2\over k_{\perp}^2}\left(1+{\omega _{pe}^2 \over
k_{\perp}^2c^2}\right)^{-1}\right]^{-1}\\ &\times\left\{
\int\vec{k}\cdot{\partial f_i^{\prime}\over\partial\vec{v_i}}\delta
^3\left( \Omega _{LH}-\vec{k}\cdot\vec{v_i}\right)d^3\vec{v_i} +
{\omega _{pe}^2\over \omega _{pi}^2}\left(1+{\omega
_{pe}^2/k^2c^2}^{-1}\right){\partial f_e^{\prime}\over\partial
v_{||}}|_{v_{||}=\Omega _{LH}/k_{||}}\right\}.
\end{eqnarray}
The first term gives wave growth/damping by the suprathermal ions
and the second denotes that by suprathermal electrons moving along
magnetic field lines. The growth rate due to a Maxwellian ion
distribution with temperature $T$ moving with bulk velocity
$\vec{U}$ may be evaluated (dropping the term in $\omega ^2/\Omega
_e^2$ in the perpendicular electron term);
\begin{eqnarray}
\nonumber\gamma _i=&{\pi\over 2}\Omega _{LH}^3{n_i^{\prime}\over
n_i}\left(1+{\omega _{pe}^2\over \omega _{pi}^2} \cos
^2\theta\right)^{-1}{1\over k^2}\int\vec{k}\cdot\left(\vec{v_i}
-\vec{U}\right){m_i\over k_BT} {f_i^{\prime}\over
n_i^{\prime}}\delta\left(\Omega
_{LH}-\vec{k}\cdot\vec{v_i}\right)d^3\vec{v_i}\\ =&{1\over
2}\sqrt{\pi\over 2}{n^{\prime}\over n_i} \left(\Omega _{LH}\over
k\sqrt{k_BT/m_i}\right)^3 \left(1+{\omega _{pe}^2\over \omega
_{pi}^2} \cos
^2\theta\right)^{-1}\left(\vec{k}\cdot\vec{U}-\Omega_{LH}\right)
\exp\left( -\left(\Omega _{LH}-\vec{k}\cdot\vec{U}\right)^2
\over\left(2k_BT/m_i\right)k^2\right) .
\end{eqnarray}
The growth rate is maximized for $\left(\vec{k}\cdot\vec{U}
-\Omega_{LH}\right)=k\sqrt{k_BT/m_i}$, hence
\begin{equation}
\gamma _{i,max}={1\over 2}\sqrt{\pi\over 2}\exp\left(-{1\over
2}\right){n_i^{\prime}\over n_i} {\Omega _{LH}\over\left(1+{\omega
_{pe}^2\over\omega _{pi}^2} \cos ^2\theta\right)} \left(\Omega
_{LH}\over k\sqrt{k_BT/m_i}\right)^2,
\end{equation}
which gives strongest wave growth at the smallest possible
wavevector, $k_{min}=\omega _{pi}/\sqrt{k_BT/m_i}$. Hence
\begin{equation}\label{eqn6}
\gamma _{i,max}=0.38{\Omega _e^2\over\omega
_{pe}^2}{n_i^{\prime}\over n_i} {\Omega _{LH}\over\left(1+{\omega
_{pe}^2\over\omega _{pi}^2} \cos ^2\theta\right)},
\end{equation}
and can be seen to be largest for $\cos\theta < \omega
_{pi}/\omega _{pe}$, in qualitative agreement with the reactive
growth rate. However the kinetic growth rate is smaller by a
factor $\sim\Omega _e^2/\omega _{pe}^2$. If the reflected ions
have a larger temperature $T^{\prime}$ than the ambient ion
temperature $T$, the growth rate is smaller still by a factor
$\sim T/T^{\prime}$. Returning to equation(\ref{eqn5}) the
accelerated electron distribution function is given (with the
assumption of steady state conditions, i.e. $\gamma =0$) by
\begin{equation}\label{eqn7}
{\omega _{pe}^2\over\omega _{pi}^2} \left(1+{\omega
_{pe}^2/k^2c^2}\right)^{-1}{\partial f_e^{\prime}\over\partial
v_{||}}|_{v_{||}=\Omega _{LH}/k_{||}
}=-{n_i^{\prime}\over\sqrt{2\pi }}{m_i\over k_BT}
\exp\left(-{1\over 2}\right).
\end{equation}
With some manipulation $\omega _{pe}^2/k^2c^2=\Omega
_{LH}^2/k^2v_{Ai}^2= \Omega _{LH}^2\cos ^2\theta
/k_{||}^2v_{Ai}^2=v_{||}^2\cos ^2\theta /v_{Ai}^2$, allowing us to
integrate to get $f_e$;
\begin{equation}\label{eqn8}
f_e^{\prime}\left(v_{||}\right)={n_i^{\prime}\over\sqrt{2\pi }k_BT/m_i}
\exp\left(-{1\over2}\right){\omega _{pi}^2\over\omega _{pe}^2}
\left(v_m-v_{||} +{\left(v_m^3-v_{||}^3\right)\cos ^2\theta\over 3v_{Ai}^2}\right)
\end{equation}
where $v_{Ae} = v_{Ai}\sqrt{m_in_i/m_en_e}$ is the electron
Alfv\'en velocity, $v_m$ is a constant of integration, the maximum velocity where
$f\left(v_m\right)=0$, and $\theta $ is the angle between the magnetic
field direction and the wavevector where the growth rate is a
maximum. We may integrate over $f_e^{\prime}$ in the range
$-v_m\rightarrow +v_m$ to obtain an expression for the number
density of energized electrons, $n_e^{\prime}$. The result is
\begin{equation}\label{eqn9}
n_e^{\prime}={n_i^{\prime}\over\sqrt{2\pi }k_BT/m_i}
\exp\left(-{1\over 2}\right){\omega _{pi}^2\over\omega _{pe}^2}
\left(v_m^2 +{v_m^4\cos ^2\theta\over 2v_{Ai}^2}\right).
\end{equation}
\citet{vaisberg83} appear to make the approximation $\left(1+\omega _{pe}^2/k^2c^2\right)^{-1}
\simeq 1-\omega _{pe}^2/k^2c^2$ to derive a slightly different form for the accelerated
electron distribution, where $v_{Ae}$ becomes a firm upper limit to $v_{||}$.
\subsection{Group Velocity}
From equation (A4) with $\omega _{pe} >>\Omega _e$
\begin{equation}
\omega ={\Omega _e\over\omega _{pe}k_{\perp }}\sqrt{\omega
_{pi}^2\left(k_{\perp }^2+ k_{||}^2\right)+\omega
_{pe}^2k_{||}^2}.
\end{equation}
Hence the wave group velocity away from the shock front is
\begin{equation}
v_g=\left|\partial\omega\over\partial k_{\perp
}\right|={\omega\over k_{\perp }} \left[{\omega
_{pe}^2k_{||}^2+\omega _{pi}^2k_{||}^2\over\omega
_{pi}^2\left(k_{\perp }^2+ k_{||}^2\right)+\omega
_{pe}^2k_{||}^2}\right].
\end{equation}
If $\omega /k_{\perp}\sim U\sim 2v_s$ then $v_g\ge v_s$ for
$k_{||}/k_{\perp}\sim\cos\theta\sim \omega _{pi}/\omega _{pe}$.
Waves excited with lower $\cos\theta $ have lower group
velocities. Since the reflected ions only move a distance
downstream on the order of a gyroradius, growth rates need to be
several times $\Omega _i$ if significant wave growth for these
waves is to occur before they are overrun by the shock. Waves at
$\cos\theta\sim\omega _{pi}/\omega _{pe}$ and above can stay ahead
of the shock and can hence grow to large intensities, even if
the growth rate is small. Hence in equation (A17) we simply put
$\cos\theta = \omega _{pi}/\omega _{pe}$ and proceed with the
calculations of the bremsstrahlung continuum. Similar conclusions result from the
use of the dispersion relation with electromagnetic and thermal corrections
\citep{mcclements97}, and will be discussed in more detail in future work.

\subsection{Applicability Criteria and Connection to Electron Acoustic Waves}
Equation (\ref{eqn8}) for the accelerated electron distribution
assumes waves are generated at a pitch angle to the magnetic field
direction given by $\cos\theta =\omega _{pi}/\omega _{pe}$ only,
yielding a maximum electron velocity of $v_{Ae}$. In realistic
conditions both the reactive and kinetic instabilities will excite
waves with pitch angles down to $\cos\theta =0$. Consequently
waves with phase speeds along the magnetic field, $\Omega
_{LH}/k_{||}$, up to $c$ will exist, allowing some electrons to be
accelerated up to relativistic energies. However a break in the
electron spectrum should still exist at the electron Alfv\'en
speed, $v_{Ae}$.

In the lower hybrid wave the damping of the ion oscillation by
electron screening is inhibited since the electrons are
magnetized. This suggests a criterion for the existence of lower
hybrid waves of electron gyroradius $<<$ wavelength, or
$v_e/\Omega _e << v_i/\omega _{pi}$. This can be reexpressed as
$T_e<< {m_e\over m_i}T_i \Omega _e^2/\omega _{pi}^2$. Assuming
$T_i={3\over 32}m_iv_s^2/k_B$ for a 4000 km s$^{-1}$ shock in Cas
A (assuming equipartition between ions and magnetic field)
gives $T_e<<3\times 10^6$ K. On these grounds we might expect
lower hybrid wave acceleration to have ceases in Cas A, since the
electrons have been heated by collisional processes to $T_e\sim
4\times 10^7$K. On the other hand this criterion appears not to be
met for the plasma conditions near the nucleus of comet Halley,
where lower hybrid waves and their associated electron
distribution function were in fact observed in situ in 1986
\citep{gringauz86,klimov86}, suggesting possibly that it is
somewhat naive. From the condition that the electron Landau
damping rate be $<<\Omega _{LH}$ we derive $k_BT_e<<m_e\Omega
_{LH}^2/k_{||}^2$, which is equivalent to the first inequality if
$k_{||}/k =\cos\theta = \omega _{pi}/\omega _{pe}$, but extends to
higher electron temperatures for $\cos\theta\rightarrow 0$, which
would be more consistent with observations. \citet{shapiro99},
motivated in part by the discovery of cometary x-ray emission,
\citep[c.f.][]{bingham97} have recently demonstrated by
particle-in-cell simulations that the observed wave activity at
comet Halley is indeed consistent with the observed electron
distribution function.

If in equation (\ref{eqn1}) we put $\phi\left(x\right)\rightarrow
2x^2-4x^4/3+\dots$ appropriate for hot ions, then with $\Omega _e
>>\omega _{pe}$
\begin{equation}\label{eqn10}
 K^L=1+{1\over k^2\lambda _{di}^2}-
 {\omega _{pe}^2\over\omega ^2}{k_{||}^2\over k^2} =0
\end{equation}
which is equivalent to equation (3.11) of \citet{begelman88},
their dispersion relation for electron acoustic waves, in the
limit of zero ion drift velocity. These authors derived their
dispersion relations from fluid equations, assuming the electrons
to be strongly magnetized, i.e. $\Omega _e > \omega _{pe}$. This requirement is not
necessary here. As discussed by \citet{begelman88}
these electron acoustic waves have very similar properties to
lower hybrid waves. The growth rate for these waves due to a
drifting Maxwellian ion distribution may be evaluated in the same
way (putting $\omega _{pi}\rightarrow 0$ and $\omega _{pi}^{\prime}\rightarrow
\omega _{pi}$ in the forgoing) with the result
\begin{equation}
\gamma _{i,max}={1\over 2}\sqrt{\pi\over 2}\exp\left(-{1\over
2}\right){\omega\omega _{pi}^2\over\omega
_{pe}^2\cos ^2\theta} \left(\omega \over
k\sqrt{k_BT/m_i}\right)^2.
\end{equation}

\clearpage

\figcaption[casafek.ps]{Data and fit for the spectral region around the Fe K
feature at 6.7 keV taken from the BeppoSAX MECS spectrum. The temperature
fitted to the thermal bremsstrahlung continuum is 3.3 keV. \label{fig1}}

\figcaption[casafig2.ps]{Data from the BeppoSAX MECS and PDS instruments, with
model bremsstrahlung spectra for a fully ionized pure oxygen plasma. The
softest model is pure thermal bremsstrahlung at 3.3 keV temperature. The harder
spectra come from electron distribution functions appropriate to lower hybrid
wave energization, for electron Alfv\'en velocities of 32, 40, and 48
atomic units. The fraction of electrons in the non-thermal distribution is
4\%.\label{fig2}}

\figcaption[casafig3.ps]{Same data as for figure 2, but with model spectra for
pure thermal and nonthermal bremsstrahlung corresponding to an electron
Alfv\'en velocities of 56, 68, and 80 atomic units, for fully ionized helium.\label{fig3}}

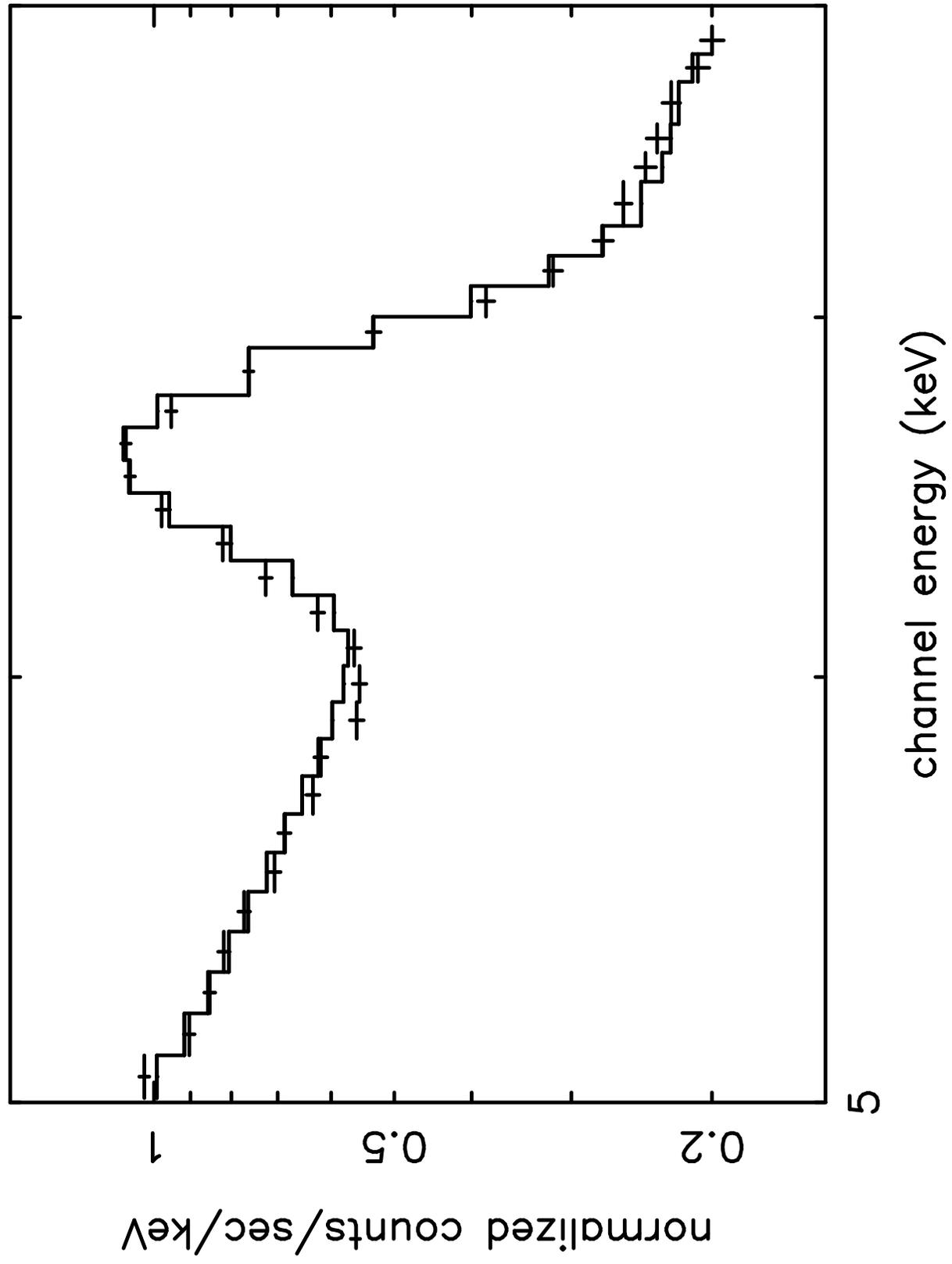
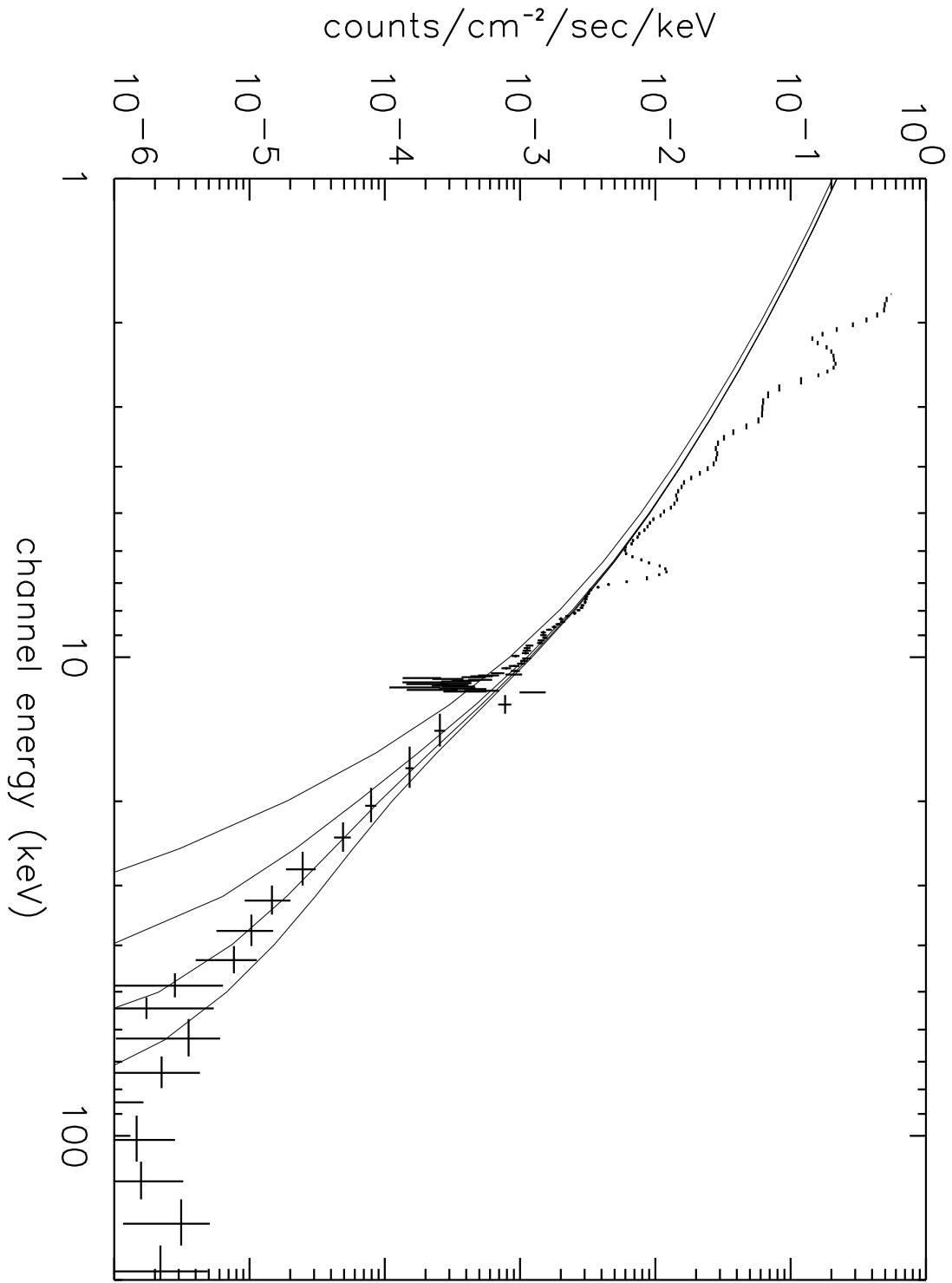
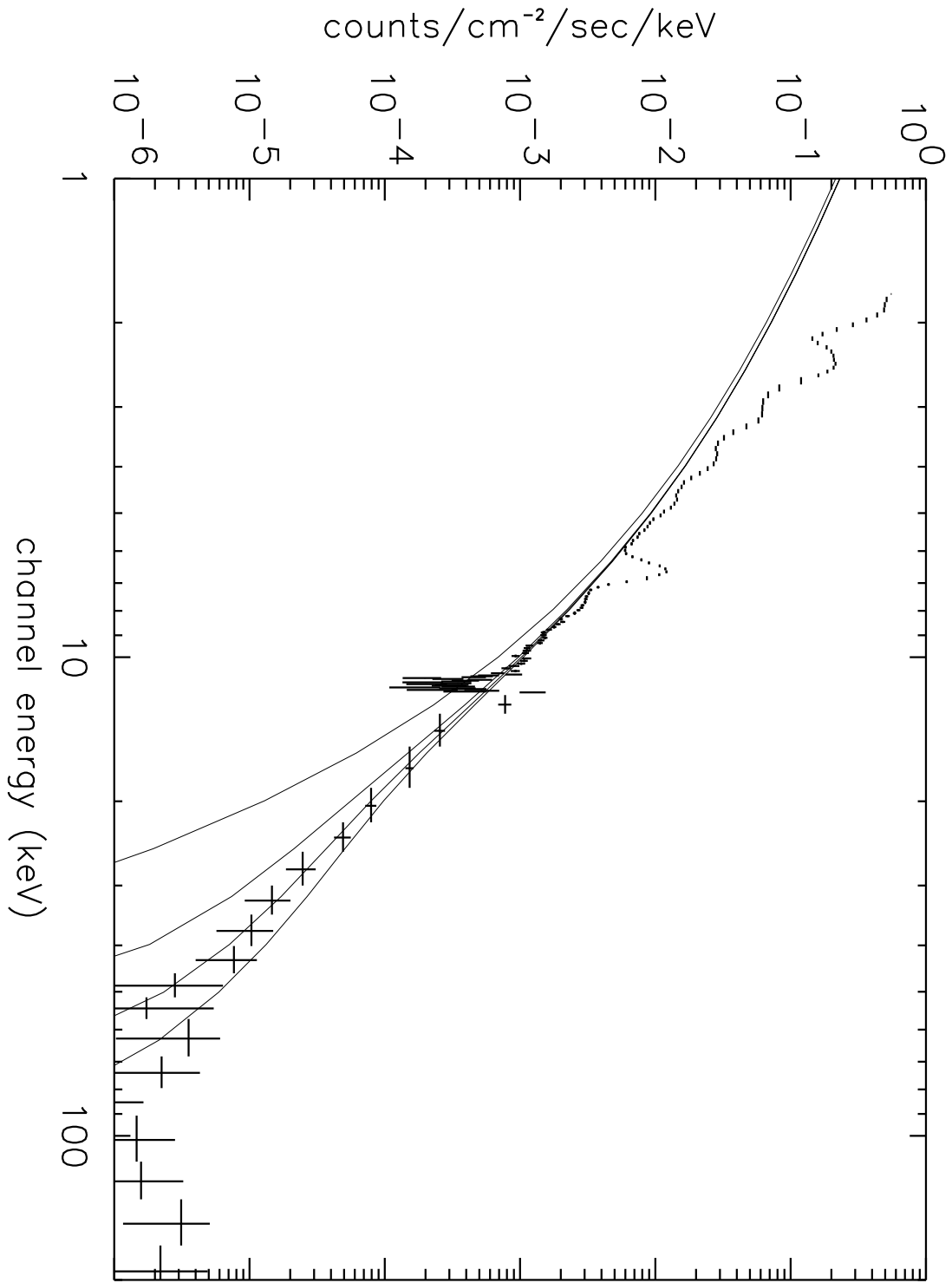
\end{document}